\documentclass[conference,compsoc]{IEEEtran}

\ifCLASSOPTIONcompsoc
  \usepackage[nocompress]{cite}
\else
  \usepackage{cite}
\fi

\ifCLASSINFOpdf
  \usepackage[pdftex]{graphicx}
\else
\fi

\usepackage{amsmath}

\begin{document}

\title{Online Learning of Temporal Dependencies \\for the Sustainable Foraging Problem}

\author{\IEEEauthorblockN{John Payne}
\IEEEauthorblockA{Department of Computer Science\\
University of Exeter\\
Exeter, UK\\
Email: jhp206@exeter.ac.uk}
\and
\IEEEauthorblockN{Aishwaryaprajna}
\IEEEauthorblockA{Department of Computer Science\\
University of Exeter\\
Exeter, UK\\
Email: aishwaryaprajna@exeter.ac.uk}
\and
\IEEEauthorblockN{Peter R. Lewis}
\IEEEauthorblockA{Faculty of Business and IT, \\ Ontario Tech University\\ Oshawa, Canada\\
Email: peter.lewis@ontariotechu.ca }}

\maketitle

\begin{abstract}
The sustainable foraging problem is a dynamic environment testbed for exploring the forms of agent cognition in dealing with social dilemmas in a multi-agent setting. The agents need to resist the temptation of individual rewards through foraging and choose the collective long-term goal of sustainability. We investigate methods of online learning in Neuro-Evolution and Deep Recurrent Q-Networks to enable agents to attempt the problem one-shot as is often required by wicked social problems. We further explore if learning temporal dependencies with Long Short-Term Memory may be able to aid the agents in developing sustainable foraging strategies in the long term. It was found that the integration of Long Short-Term Memory assisted agents in developing sustainable strategies for a single agent, however failed to assist agents in managing the social dilemma that arises in the multi-agent scenario.
\end{abstract}

\IEEEpeerreviewmaketitle

\section{Introduction}
Episodic reinforcement learning \cite{gronauer2022multi} and neuro-evolution (NE) algorithms \cite{papavasileiou2021systematic} have been widely studied for multi-agent systems with cooperative task requirements. Multi-agent foraging problem is such a testbed \cite{zedadra2017multi} where the impact of agent-agent interactions in an episode can be used by the agents to decide on an optimal policy to cooperate for task success in the following episodes. The agent tasks in foraging problems often involve environment exploration as well as identification, transport and sharing of resources which can be achieved through episodic learning. However, considering a long-term goal like the sustainability of resources in the foraging problem using episodic learning requires agents to die again and again to possibly identify unsustainable actions at every episode\cite{aishwaryaprajna2023exploring}.

Climate change and sustainable resource management are often considered wicked social problems, the solutions to which require one-shot operations without the opportunity to learn by trial and error \cite{rittel1973dilemmas}. This is because, unlike tame problems where a variety of solutions can be attempted without consequence, any attempted solution for a wicked problem has irreversible consequences. For example in the context of the real-world sustainability of a resource, once the resource has been overused or exhausted it is not possible to go back and attempt to use it more sustainably. Likewise, a possible solution to the sustainable foraging problem \cite{aish2023sustainable} could allow the agents to learn a sustainable strategy of resource collection in a one-shot manner within a single lifetime rather than with several episodes of resource depletion and agent deaths. This paper explores online learning methods to investigate if agents are able to resist the individual temptations of maximising resource collection and achieve the long-term collective goal of sustainability.

Extensive research has been conducted on the social dilemmas present in multi-agent systems, where a social dilemma arises when there is tension between individual and collective goals. An example of this is the Tragedy of Commons where individual agents with open access to a shared resource independently adopt selfish strategies contrary to cooperative strategies that serve the common good of the agents, leading to depletion of the common resource. However, much of this work on multi-agent social dilemmas involved simplified environments within the context of game theory and is also focused on agent-agent interactions relating to the social dilemma rather than the long-term implications of the agent-environment interactions. 

Episodic learning is not enough for one-shot problems or those where agents need to adapt to a dynamic environment or the actions of other agents. The Open Racing Car Simulator (TORCS) is such a problem, where offline controllers were found to be too predictable and unable to adapt to varying environments \cite{cardamone2009NE}. An approach of neuro-evolution with augmenting topologies (NEAT) \cite{whiteson2006evolutionary} to both evolve a fast controller from scratch and optimise an existing controller for a new track using online learning was discussed.

The sustainable foraging problem can be considered a Partially Observable Markov Decision Process (POMDP) where agents only have partial observability of the environment. There has been significant research in exploring the inclusion of Long Short-Term Memory (LSTM) in a neural network to compensate for partial observability and capture long-term dependencies in sequences of observations \cite{hausknecht2015deep} \cite{zhu2017improving}. Deep Recurrent Q-Networks (DRQNs) with LSTM were shown to compensate for missing frames in an arcade game environment from historical information, where agents were only given access to a limited fraction of the frames.

In this work, we first discuss the one-shot nature of the sustainable foraging problem. This leads us to explore the differences in agent actions based on episodic and online learning algorithms for this one-shot problem. We implement an online neuro-evolution as well as an online DRQN as the agent's deliberative architecture. We then augment both online neuro-evolution and DRQN with LSTM using sequences of observations to explore whether agents are aware of the dynamic environment and can make long-term decisions.

\section{A Brief on the Sustainable Foraging Problem and its One-shot Nature}

The sustainable foraging problem has been proposed as a dynamic social environment testbed for exploring the forms of agent cognition needed to achieve sustainability \cite{aish2023sustainable}. The agents need to collect resources from the environment to gain the energy necessary for survival. The problem involves three environment types: \textit{forest}, \textit{pasture}, and \textit{desert} characterised by the replenishment rate of the resources. The rate of replenishment is directly proportional to the amount of available resources. Agents within the problem have a choice of two actions at every time step: \textit{greedy} or \textit{moderate}. Greedy agents continuously gather resources regardless of whether they are in immediate need to survive, whereas moderate agents only gather resources when they are needed for survival. Regardless of the foraging strategies adopted by agents, agents will be able to survive indefinitely in the forest environment type but they will not be able to survive in the desert environment type. The pasture environment type can support agents indefinitely provided too many resources are not removed from the environment too quickly. Agents will only be able to survive if they adopt moderate foraging as greedy strategies lead to the rapid depletion of the resources and a reduction in the replenishment rate to the extent that it will no longer be able to support agents.

The chosen environment types lie on a tradeoff spectrum based on the relationship between resource availability and the consumption rate of the agents. A compromise solution of maximising resource availability and minimising consumption to achieve sustainability is only required in the pasture environment type. To capture this tradeoff while exploring the agent-environment interactions, this paper is only focussed on the pasture environment type. It should be noted that transitions between the environment types may occur when the number of agents changes in the environment. Death, reproduction or migration can be possible causes for the change in the number of agents, but the latter two causes are not considered in this paper. 

It is assumed that the agents are successful in this problem if they can act and reason based on the environment type they are in and achieve sustainability. This means that, in the pasture environment type, the agents will be successful if they all can take moderate actions collectively and escape the imminent Tragedy of Commons. The collective non-exploitative behaviour has to be ensured before the resources reach a point of no return (an irreversible state). This one-shot nature of the sustainable foraging problem makes it challenging for the agents to reason about the depleting gradient of the resources within the right time and act accordingly. 

\section{Episodic vs. Online  Neuro-evolution }

Episodic neuro-evolution has been previously applied to the sustainable foraging problem \cite{aishwaryaprajna2023exploring}. Episodic (or offline) approaches require agents to experience several episodes of interaction with the environment where the weights of the neural network are updated at the end of each episode. The weight updates are accepted when there is an increase in the cumulative reward in comparison to the original network weights. The results from the previous study show that a deliberative agent architecture with offline neuro-evolution is not enough for the agents to ensure the sustainability of the resources. The n-player game arising from the dilemma makes individual resource collection more tempting than the collective goal of survival through sustainability. This implies that the agents have to die several times to learn the impact of their actions.

The environment is as outlined in the implementation for the pasture environment type in \cite{aishwaryaprajna2023exploring} with an initial resource level of 500 per agent and a resource growth rate of 1.005. In all of the experiments, the initial energy of an agent is chosen as 100. The agents can consume 5 units of resource at every time step at most and the cost of surviving each step is 2 units. Agents can choose moderate or greedy actions, dictated by the energy threshold, a moderate action will result in the agent only gathering when its energy level is below the threshold whereas a greedy action will result in the agent gathering regardless of its current energy level. The logarithm of the agent's energy level is used as the reward function for all implementations which results in agents attempting to maximise the amount of resources they can gather.

Results are plotted with agent count on the left axis representing the mean number of agents across independent runs, alive agents show the survival rate or proportion of surviving agents across experiments, and the proportion of agents choosing each action shown by the threshold is also indicated on the left axis. Resources are shown using the right axis to indicate the mean resource level in the environment across experiments.

\begin{figure}[!t]
\centering
\includegraphics[width=2.5in]{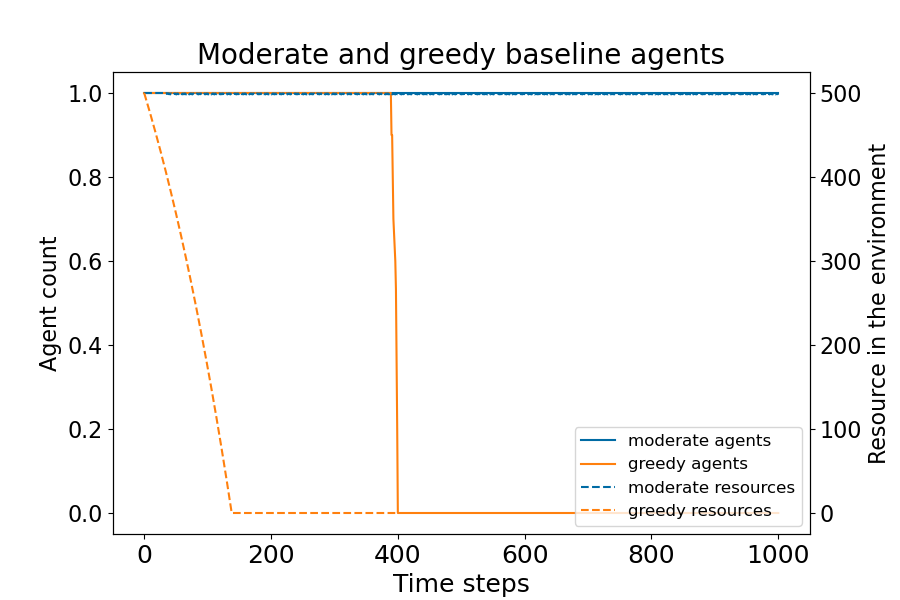}
\caption{Mean simulation results across 30 independent runs showing baseline behaviour of moderate and greedy agents}
\label{Moderate_Greedy}
\end{figure}

Baselines were obtained with moderate and greedy agents that chose the same action for every time step. The energy level threshold for the moderate action is set to 50. Figure \ref{Moderate_Greedy} shows the average behaviour from 30 independent runs of a single moderate and greedy agent over 1000 time steps. The moderate agent waits until its energy level is below the threshold before gathering and then only gathers when its energy level is below the threshold. This results in sustainable use of the resource where the agent can balance its energy needs with the replenishment rate of the environment. The greedy agent agent continuously gathers resources. This results in the overuse of the resources characterised by a rapid decline in the resource level until it is depleted. At this point, the agent is no longer able to obtain any resources from the environment as there is no longer any replenishment of the resources, the agent's energy level then inevitably declines until the agent dies.

We implement a method of online neuro-evolution similar to \cite{cardamone2010learning}, where the networks are updated at each time step rather than at the end of an episode. In general, neuro-evolution can be used to evolve the weights or topology of a network or both, here we have only evolved the weights of the network. The network used by the agents consists of a two-layer fully connected neural network of three input neurons corresponding to the state parameters: the resource level in the environment, the number of alive agents, and the number of resource collecting agents, as well as an additional input neuron for each choice of energy threshold available to the agent corresponding to the number of agents that chose each respective threshold during the last time step. The second layer in the network is a hidden layer of three hidden neurons, with a Rectified Linear Unit (ReLU) activation function to introduce non-linearity into the model. The output layer consists of one neuron per choice of energy threshold available to the agents with a softmax activation function which transforms the outputs into a probability distribution over the actions. Each action corresponds to an energy threshold which determines whether the action is greedy or moderate.

Each agent independently maintains a population of 30 neural networks, initialised with random weights and a fitness value based on the agent's initial energy level. The network used to decide an action for the agent is obtained from a softmax distribution of the fitness values of the networks, where a network with a higher fitness has a higher probability of being chosen. In each time step the observation is passed to the selected network to obtain the decided action and at the end of the time step the fitness of the selected network is updated with the agent's reward. Two parent networks are then obtained by tournament selection of 5 networks in the population and used to produce a new network via arithmetic crossover of the network weights. The new network weights are mutated with a rate of 0.2 by applying Gaussian noise with a scale of 0.06, before replacing the network with the lowest fitness in the population. The stated NE parameters were chosen to maximise agent reward during tuning.

\begin{figure}[!t]
\centering
\includegraphics[width=2.5in]{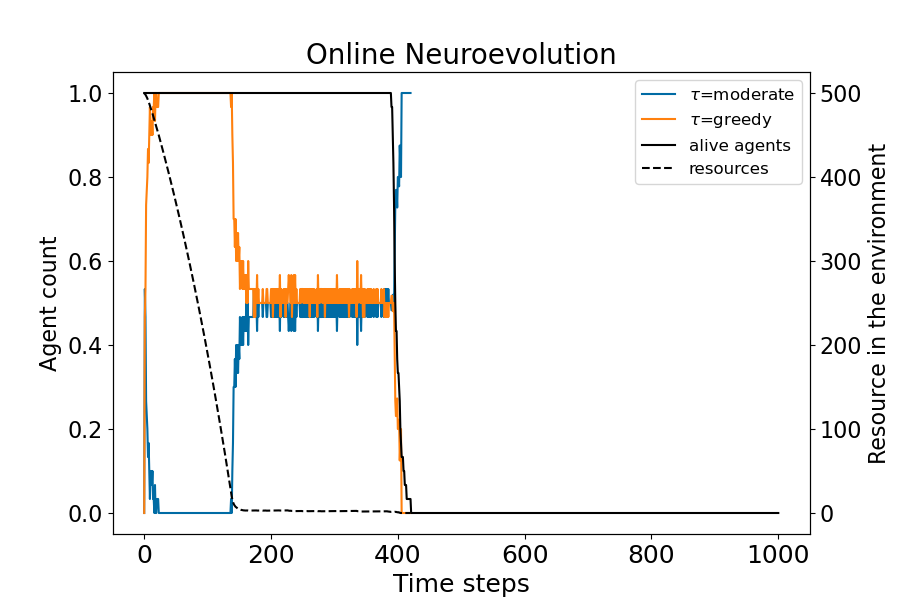}
\caption{Mean simulation results for a single online neuro-evolution agent with a choice of greedy or moderate actions, averaged over 30 independent runs of 1000 time steps each.}
\label{Online_NE1}
\end{figure}

For the case of a single agent using online NE, illustrated in figure \ref{Online_NE1}, agents start with random actions and rapidly learn a greedy strategy resulting in the resource depletion and death of agents typical in the case of greedy agents. Once the resources have been depleted the agent makes more use of the moderate action in an attempt to gain more reward however this is too late to help the agent due to the lack of replenishment once the resource is depleted. It is interesting to refer to the case of a single agent with episodic NE, that the agent becomes moderate after a few initial episodes of dying and stays moderate for the rest of the episodes \cite{aishwaryaprajna2023exploring}.

\begin{figure}[!t]
\centering
\includegraphics[width=2.5in]{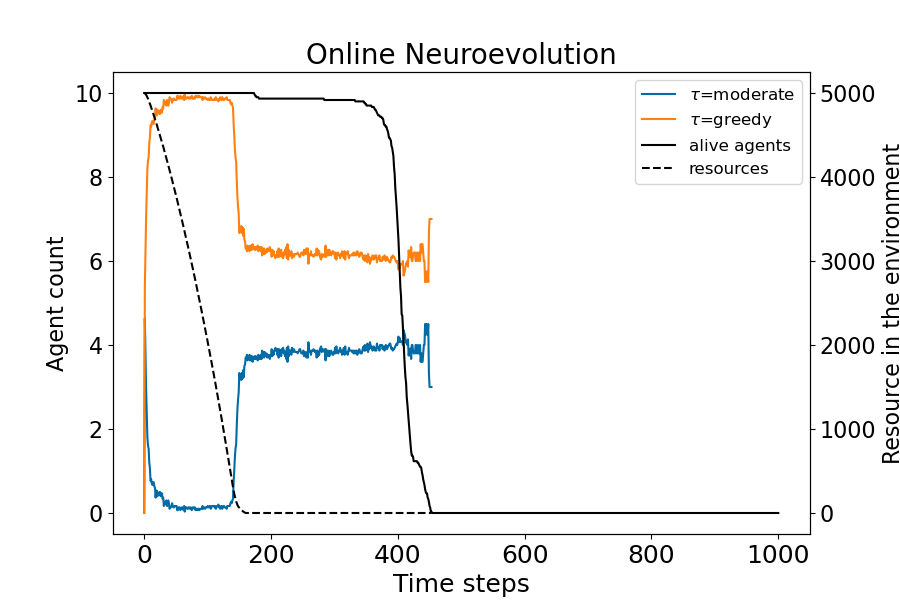}
\caption{Mean simulation results for 10 online neuro-evolution agents with a choice of greedy or moderate actions, averaged over 30 independent runs of 1000 time steps each.}
\label{Online_NE10}
\end{figure}

For the case of 10 agents with online NE, illustrated in figure \ref{Online_NE10}, agents initially choose random actions but all quickly learn greedy strategies in an attempt to maximise their cumulative reward. This leads to rapid overuse of the resource resulting in its rapid depletion and the Tragedy of Commons. Once the resource has been depleted, agents can no longer increase their energy level and attempt a mix of actions until their reserves are expended and all agents die. This replicates the results obtained in the previous offline neuro-evolution approach \cite{aishwaryaprajna2023exploring} whilst maintaining a one-shot approach to the sustainable foraging problem. Whilst the online neuro-evolution agents were not able to find sustainable strategies, we would not expect them to as they only have knowledge of the current time step and incentive to maximise their immediate reward. However, these results show that online neuro-evolution agents can learn the greedy strategy we expect of them within a single lifetime, which is a requirement of wicked social problems.

\begin{figure}[!t]
\centering
\includegraphics[width=2.5in]{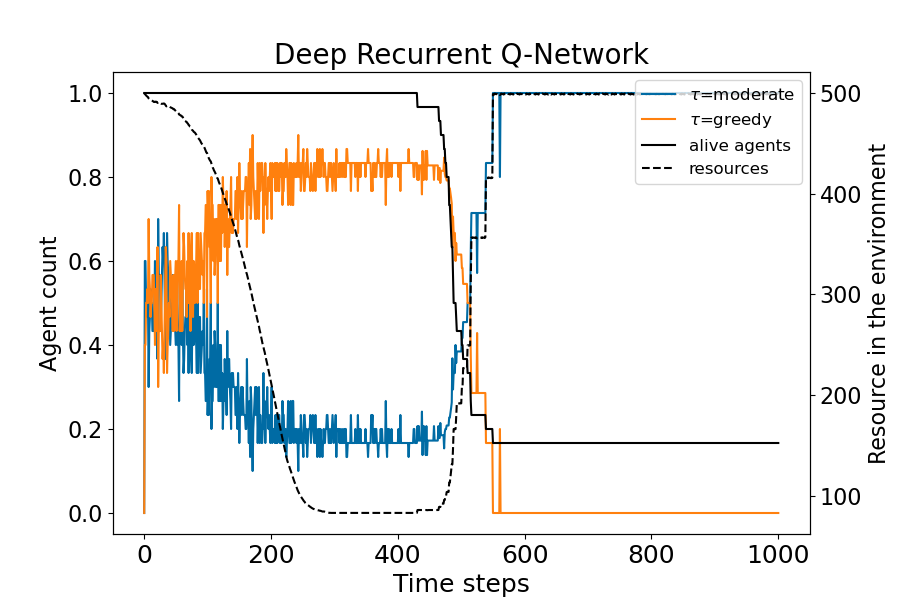}
\caption{Mean simulation results for a single DRQN agent with a choice of greedy or moderate actions, averaged over 30 independent runs of 1000 time steps each.}
\label{DRQN_1}
\end{figure}

\begin{figure}[!t]
\centering
\includegraphics[width=2.5in]{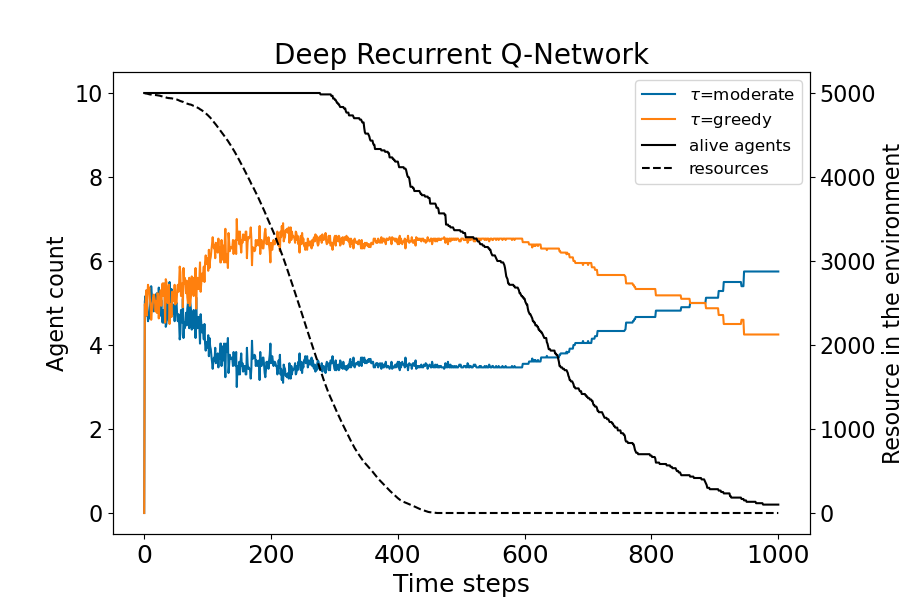}
\caption{Mean simulation results for 10 DRQN agents with a choice of greedy or moderate actions, averaged over 30 independent runs of 1000 time steps each.}
\label{DRQN_10}
\end{figure}

Simulations were repeated with the implementation of DRQN given the partially observable nature of this problem. Each agent has the same topology, activation functions, and initialisation as the online NE, where the network weights were instead updated using Q-learning where the quality of the solution was dictated by the agent's reward, as in the online neuro-evolution case network update was applied at each time step. The results for this implementation are shown in figure \ref{DRQN_1}, where we see that after starting the initial time step with random actions the DRQN agents migrate towards a greedy strategy. We also see that the DRQN agents take longer to develop the expected strategy and as a result, a `lucky' 16\% of the agents survive all 1000 time steps by failing to learn a greedy strategy. Similar to the case of the online neuro-evolution agents, we see in figure \ref{DRQN_10} that for the case of 10 DRQN agents, agents again take longer to adopt greedy strategies but ultimately the majority do so leading to the depletion of the resources and death of all agents.

\section{Can LSTM Provide Insights into the Negative Gradient of Resources?}

To explore if learning temporal dependencies allows the agents to make better decisions, we replace the first fully connected layer in the agent's deliberative network with a single-layer LSTM with the same number of nodes. Sigmoid and tanh activation functions are used within the LSTM layer, and again a softmax activation function transforms the outputs into a probability distribution over the actions. The addition of LSTM allowed agents to be provided with a rolling sequence of the last 25 observations and consider how the environment may be changing in response to their actions. Agents were given a choice of three moderate actions with energy thresholds of 30, 50, and 80 respectively and a greedy action represented by an energy threshold of 50000. The energy threshold of 50000 is larger than the maximum energy an agent can obtain during the entire simulation run with 3000 time steps and therefore always corresponds to a greedy action.

\begin{figure}[!t]
\centering
\includegraphics[width=2.5in]{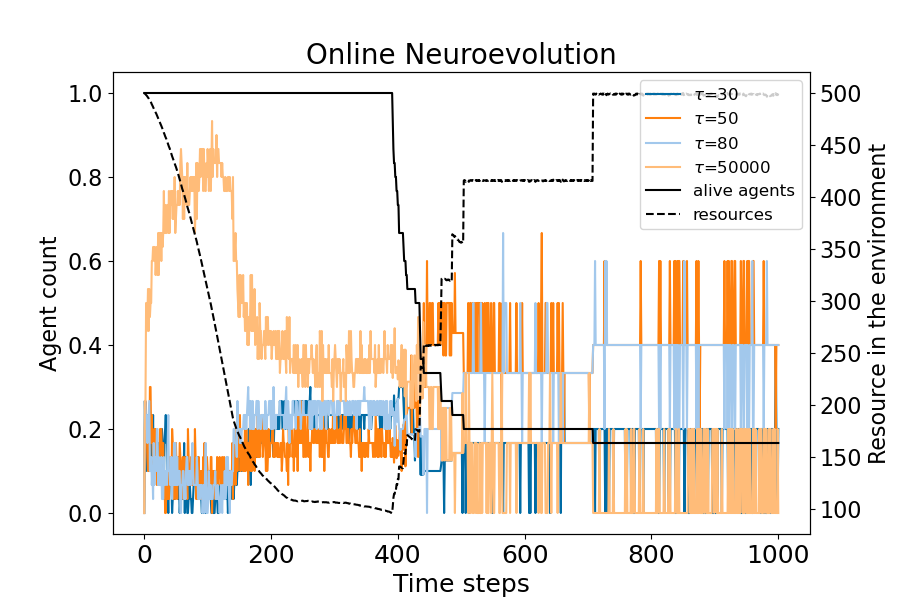}
\caption{Mean simulation results for a single online neuro-evolution agent with a choice of a range of actions from specified thresholds, averaged over 30 independent runs of 1000 time steps each.}
\label{NE_1Agent}
\end{figure}

\begin{figure}[!t]
\centering
\includegraphics[width=2.5in]{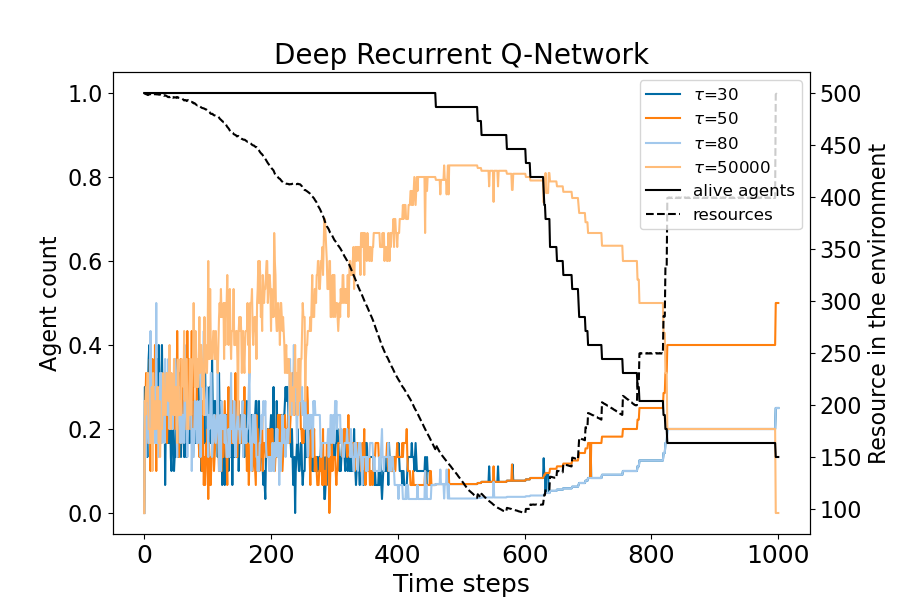}
\caption{Mean simulation results for a single DRQN agent with a choice of a range of actions from specified thresholds, averaged over 30 independent runs of 1000 time steps each.}
\label{DRQN_1Agent}
\end{figure}

Figure \ref{NE_1Agent} illustrates the case of a single online neuro-evolution agent without LSTM. The addition of multiple thresholds for the agent to choose from reduces the speed at which the agents can determine the action that maximises their reward however the vast majority of agents still rapidly adopt greedy strategies leading to the severe depletion of the resource and ultimately the death of agents. Once resources are depleted the number of greedy agents drops sharply as agents try to find more moderate actions however as the resource is already depleted this does not help and all greedy agents die leaving only the `lucky' minority of agents that failed to maximise their reward. Figure \ref{DRQN_1Agent} shows the behaviour of the DRQN agent for the same scenario, here we can see that the DRQN agent takes much longer than the neuro-evolution agent to learn a greedy strategy, potentially indicating it is less able to handle the larger action space resulting from a range of thresholds. DRQN agents in this case are unable to change their behaviour before the resources are depleted and all agents that were initially able to learn a greedy strategy then died.

\begin{figure}[!t]
\centering
\includegraphics[width=2.5in]{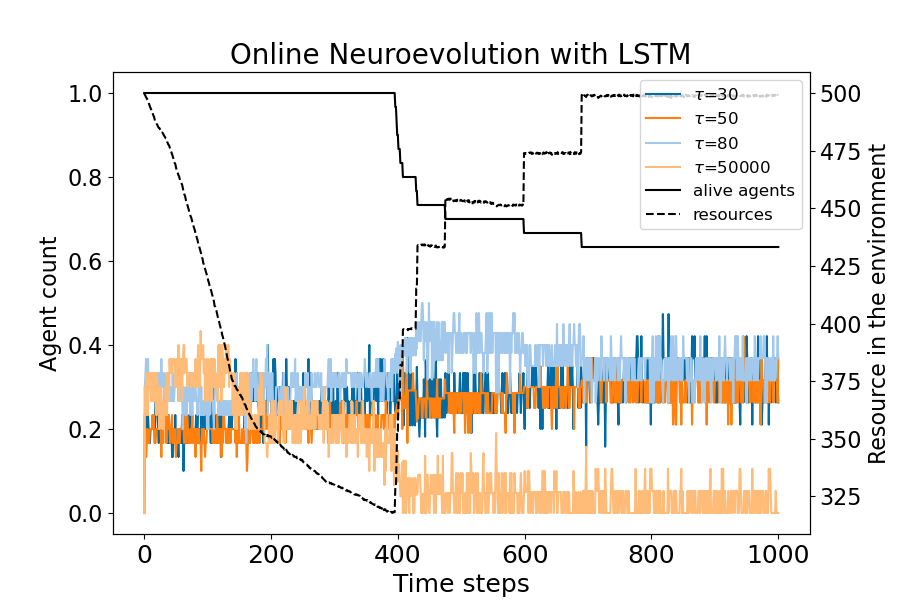}
\caption{Mean simulation results for a single online neuro-evolution agent with LSTM and a choice of a range of actions from specified thresholds, averaged over 30 independent runs of 1000 time steps each.}
\label{LSTM_NE_1Agent}
\end{figure}

\begin{figure}[!t]
\centering
\includegraphics[width=2.5in]{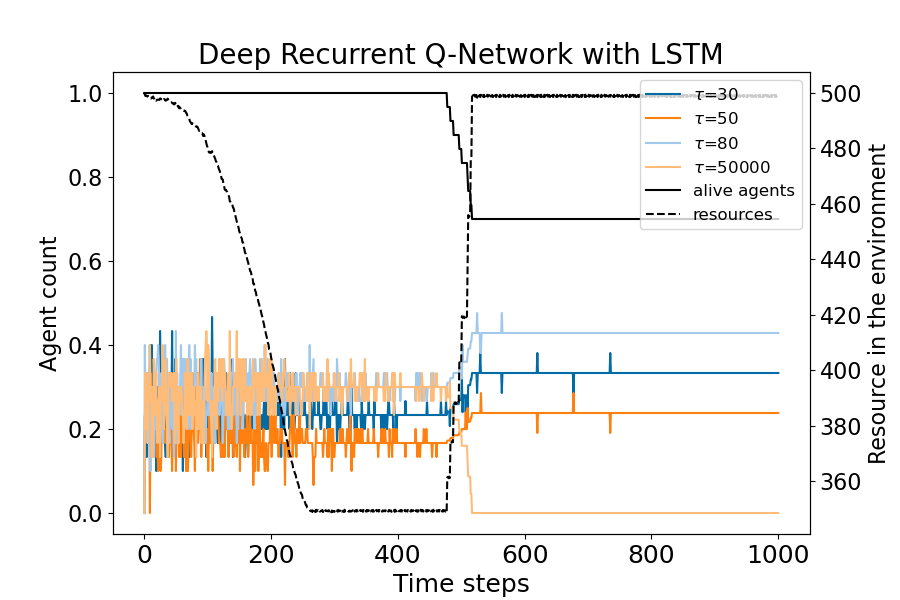}
\caption{Mean simulation results for a single DRQN agent with LSTM and a choice of a range of actions from specified thresholds, averaged over 30 independent runs of 1000 time steps each.}
\label{LSTM-DRQN_1Agent}
\end{figure}

For a single agent, the addition of LSTM in the neural network demonstrates significantly different behaviour. A minority of agents continue to prioritise immediate rewards resulting in death however the distribution of agent decisions is more mixed and agents are often able to develop sustainable strategies. Figure \ref{LSTM_NE_1Agent} illustrates online neuro-evolution with LSTM where the agent initially prefers the greedy action, however, when the resources in the environment start depleting rapidly the agent learns to reduce greedy actions stabilising the resource gradient. The agent then adopts a mix of greedy and moderate actions that increase their reward without depleting resources. The addition of LSTM to the DRQN agent, figure \ref{LSTM-DRQN_1Agent}, demonstrates the agent is less eager to adopt very greedy strategies and in most of the runs adopts moderate strategies with the awareness of the depleting gradient of resources.

\begin{figure}[!t]
\centering
\includegraphics[width=2.5in]{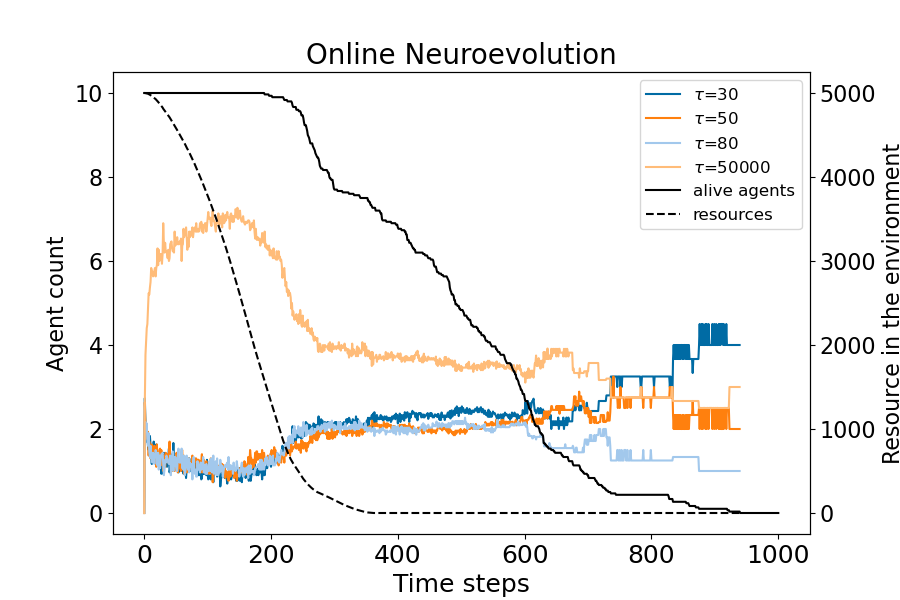}
\caption{Mean simulation results for 10 simultaneous online neuro-evolution agents and a choice of a range of actions from specified thresholds, averaged over 30 independent runs of 1000 time steps each.}
\label{NE_10Agents}
\end{figure}

\begin{figure}[!t]
\centering
\includegraphics[width=2.5in]{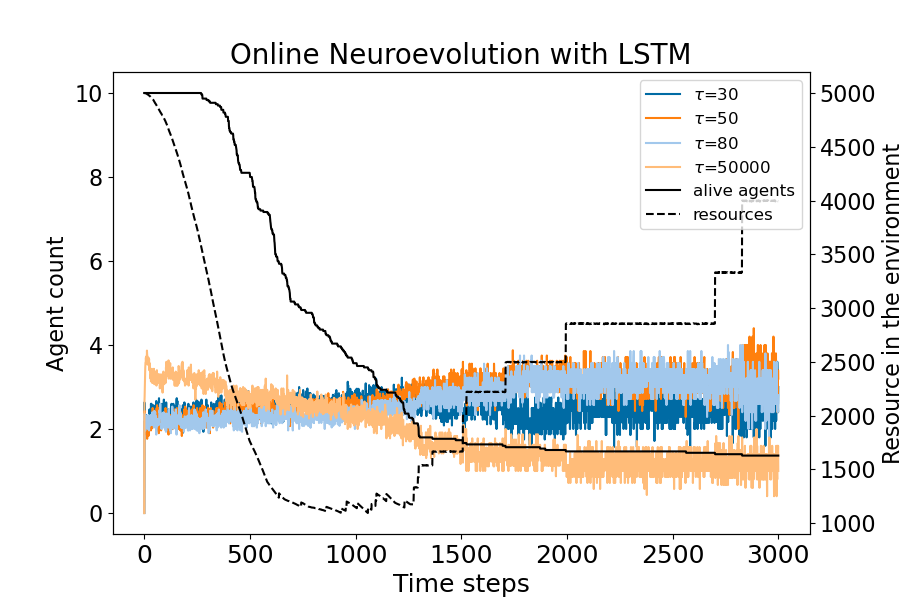}
\caption{Mean simulation results for 10 simultaneous online neuro-evolution agents with LSTM and a choice of a range of actions from specified thresholds, averaged over 30 independent runs of 3000 time steps each.}
\label{LSTM_NE_10Agents}
\end{figure}

In the case of 10 online neuro-evolution agents that can choose from a range of energy thresholds, illustrated in figure \ref{NE_10Agents}, agents take longer to decide on the greedy action however this is still chosen by the majority of agents early in the simulation. Once resources are depleted, the agents attempt other strategies but do not make an impact on sustainability. When agents are augmented with LSTM, figure \ref{LSTM_NE_10Agents}, the distribution of agent decisions early in the simulation is more mixed as in the single agent scenario. As the resources deplete, agents reduce their use of greedy actions which prolongs their lifetime but does not have a significant enough impact to bring about sustainability. Once resources are depleted agents begin to die, and the number of alive agents reduces steadily with agents only surviving the full 3000 time steps in 14\% of runs. The single-agent case demonstrates that the addition of LSTM enables agents to learn sustainable actions through observing of the impact their actions on the environment over time, which is not the case for the multi-agent scenario. This is because whilst the agents may be able to balance long-term and short-term rewards individually, as part of a group they fall victim to the Tragedy of Commons, where defectors from the sustainable strategy can obtain a greater reward than the cooperators.

\begin{figure}[!t]
\centering
\includegraphics[width=2.5in]{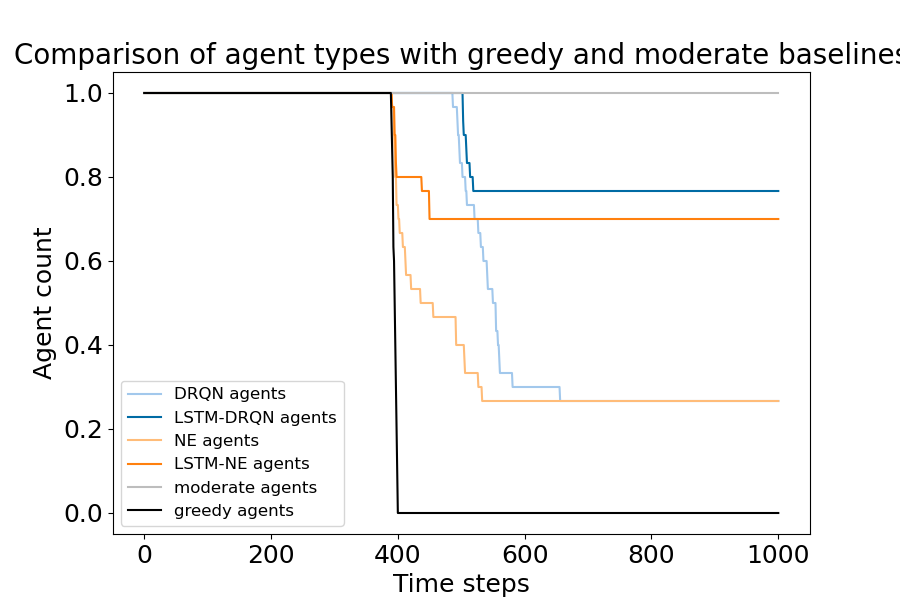}
\caption{Comparison of mean survival rate by agent type for the single agent scenario, where agents choose from a range of actions corresponding to specified thresholds. Results were obtained via 30 independent runs of 1000 time steps each for each agent type.}
\label{Model_Comp_1Agent}
\end{figure}

Figure \ref{Model_Comp_1Agent} shows a comparison of mean alive agents for a single agent of all agent types given a choice of actions from a range of moderate and greedy actions. We see that the addition of LSTM enables agents to find significantly more sustainable strategies than their non-LSTM counterparts, additionally whilst online neuro-evolution agents tend to be greedier early in the simulation they modify their strategies and recognise the impact of their actions on the environment sooner than the DRQN agents.

\begin{figure}[!t]
\centering
\includegraphics[width=2.5in]{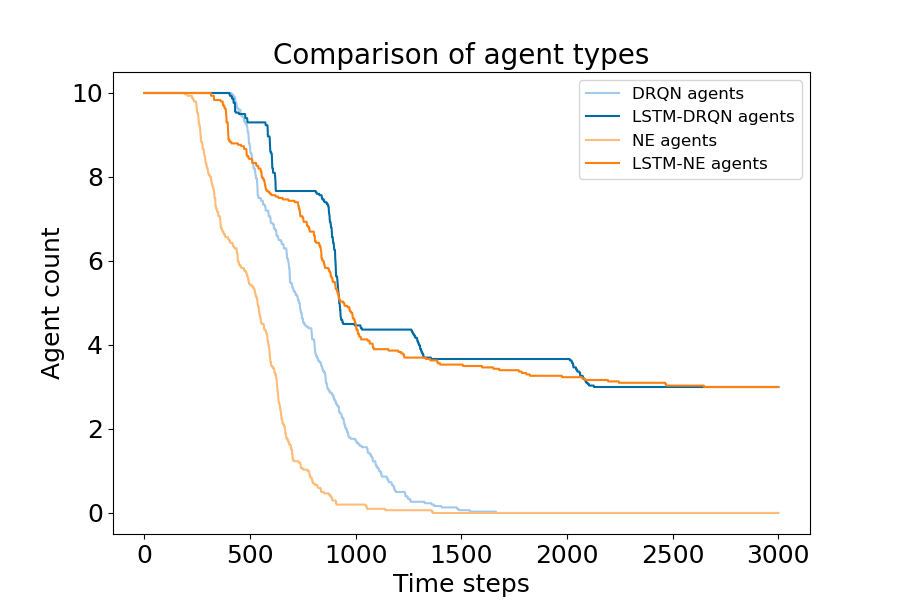}
\caption{Comparison of mean alive agents by agent type for multi-agent scenario, where agents choose from a range of actions corresponding to specified thresholds. Results were obtained via 30 independent runs of 3000 time steps for each agent type.}
\label{Model_Comp_10Agents}
\end{figure}

For 10 agents, illustrated in figure \ref{Model_Comp_10Agents}, no online neuro-evolution or DRQN agents survive as they quickly learn greedy strategies to maximise their short-term reward. The survival rate of agents with LSTM is higher than those without suggesting these agents are attempting to consider sustainability, however, all types show a declining agent count and death for the majority of agents with none managing to overcome the Tragedy of Commons.

\section{Conclusions}
In this work on the sustainable foraging problem, we first investigate online learning methods that allow agents to learn the impact of their actions within a single lifetime. An implementation of online neuro-evolution enabled agents to learn the expected strategies within one episode where many smaller, more frequent network updates are used in comparison to episodic neuro-evolution. This was compared to a DRQN implementation that showed DRQN agents are also able to learn the expected strategy within a single lifetime, however, online neuro-evolution agents were able to learn the expected greedy behaviour both more often and faster than DRQN. However, both DRQN and online neuro-evolution agents were not able to balance short-term and long-term goals and learn a sustainable strategy for either the single or 10-agent scenario, indicating that whilst online learning enables the problem to be attempted one-shot it does not aid the agents in finding new strategies other than those that maximise their immediate reward.

We also investigated the potential for LSTM to grant agents temporal awareness from historical information. Both online neuro-evolution and DRQN agents, when augmented with LSTM can make sustainable actions in a single-agent scenario. However, the temporal knowledge from LSTM is not enough to deal with the social dilemma present in the n-player game of the sustainable foraging problem. The LSTM implemented here explores temporal dependencies from sequences of observations within the network itself. Exploring temporal awareness through an explicit meta-layer to enable agents to have reflective capabilities would be an obvious next step.

\bibliographystyle{IEEEtran}
\bibliography{bibliography}

\end{document}